\documentclass[a4paper,fleqn,usenatbib]{mnras}
\usepackage{newtxtext,newtxmath}
\usepackage[T1]{fontenc}
\usepackage{ae,aecompl}
\usepackage{graphicx}	
\usepackage{amsmath}	
\usepackage{amssymb}	
\usepackage{color}      
\definecolor{darkgreen}{rgb}{0,0.5,0}
\definecolor{magen}{rgb}{0.79,0.08,0.48}

\newcommand{\grale}{\textsc{Grale}}
\newcommand{\lenstool}{\textsc{Lenstool}}
\newcommand{\glafic}{\textsc{Glafic}}
\newcommand{\as}{$^{\prime\!\prime}$}
\newcommand{\am}{$^{\prime}$}

\renewcommand{\vec}[1]{\boldsymbol{#1}} 

\title[Line of sight structure in the Hubble Frontier Field cluster MACSJ0717]
{Evidence for the line of sight structure in the Hubble Frontier Field cluster, MACSJ0717.5+3745}

\author[L.L.R. Williams et al.]{
Liliya L.R. Williams,$^{1}$\thanks{E-mail: llrw@umn.edu (LLRW)}
Kevin Sebesta,$^{1}$
Jori Liesenborgs$^{2}$
\\
$^{1}$School of Physics and Astronomy, University of Minnesota, 116 Church Street, Minneapolis, MN 55455, USA \\
$^{2}$Expertisecentrum voor Digitale Media, Universiteit Hasselt, Wetenschapspark 2, B-3590, Diepenbeek, Belgium\\
}

\date{Accepted XXX. Received YYY; in original form ZZZ}
\pubyear{2017}

\begin{document}
\label{firstpage}
\pagerange{\pageref{firstpage}--\pageref{lastpage}}
\maketitle

\begin{abstract}
MACS J0717 is the most massive and extended of the Hubble Frontier Field clusters. It is one of the more difficult clusters to model, and we argue that this is in part due to the line of sight structure (LoS) at redshifts beyond 2. We show that the \grale~mass reconstruction based on sources at $3\!\!<\!\!z_s\!\!<\!\!4.1$ has at least $10^{13}M_\odot$ more mass than that based on nearby sources, $z_s\!<\!2.6$, and attribute the excess mass to a putative LoS, which is at least $75$\as~from the cluster center. Furthermore, the lens-model fitted $z_s$'s of the recent Kawamata et al. reconstruction are biased systematically low compared to photometric $z_s$'s, and the bias is a function of images' distance from the cluster center. We argue that these mimic the effect of LoS. We conclude that even in the presence of 100-200 images, lens-model adjusted source redshifts can conceal the presence of LoS, demonstrating the existence of degeneracies between $z_s$ and (sub)structure. Also, a very good fit to image positions is not a sufficient condition for having a high fidelity mass map: Kawamata et al. obtain an rms of $0.52$\as~for 173 images of 60 sources; our \grale~reconstruction of the exact same data yields a somewhat different map, but similarly low rms, $0.62$\as. In contrast, a \grale~model that uses reasonable, but fixed $z_s$ gives a worse rms of $1.28$\as~for 44 sources with 126 images. Unaccounted for LoS can bias the mass map, affecting the magnification and luminosity function estimates of high redshift sources.
\end{abstract}


\section{Introduction}   

In the last 15 years gravitational lensing has become an indispensable tool in astrophysics. One of its most notable uses stems from its magnification property, and hence the ability of large concentrations of mass to act as cosmic telescopes.  Galaxy clusters, the subject of this paper, can magnify background sources, resulting in better angular resolution and higher observed fluxes, resulting in observations of distant faint galaxies, and estimation of their luminosity functions \citep[e.g.,][]{bou17,kar17,ala16,mcl16,bow15,fin15}.

However, because cluster optics are uneven, they have to be characterized before being used as cosmic telescopes. Characterization is achieved by means of mass reconstruction using numerous multiply imaged background sources. Since even the strongest and the most studied cluster lenses with $\mathcal{O}(100)$ images do not have enough modeling constraints to suppress all lensing degeneracies, the recovered cluster mass distributions are not unique \citep{pri17,lim16}, and the best strategy to assess systematic uncertainties is to base these on a range of reconstructions that use different lens inversion methods. This is the basic philosophy adopted by the Hubble Frontier Fields Survey (HFF; PI: J. Lotz). 

In this paper we will concentrate on one of the HFF clusters, MACSJ0717.5+3745 (hereafter MACS J0717), the largest in extent cosmic lens known. Instead of exploring the effect of degeneracies among all the models submitted in response to the STScI's call, as we did in \cite{pri17} for Abell 2744 and MACS J0416, we explore the consequences of degeneracies that exist due to the small number of available spectroscopic redshifts. While the total number of lensed sources and images in the Frontier Fields data of MACS J0717 is large, $\sim 50$ and $\sim 150$, respectively, the cluster has only 9 systems with spectroscopic redshifts.\footnote{In one additional system, \#68 in the nomenclature of \cite{lim16}, one, but not the other two, of its images has its spec-$z$ measured by Clement et al. (in prep.).} 

The importance of source redshifts for the lens reconstruction is sparsely explored in the literature. Early work \citep{kne94,bar95} suggested using cluster lensing to infer the redshift distribution (or just the most probable redshift) of background sources. However, the most common practice today is to use known source redshifts as constraints for cluster mass models. \cite{joh16} carried out a systematic exploration of the effect of spec-$z$'s using a parametrically created mock cluster, Ares \citep{men17}, and a parametric lens inversion method. Though their approach is quite different from ours---we use a real cluster and free-form lens inversion method---our conclusions are similar: spectroscopic redshifts are critical for obtaining an accurate and unbiased cluster mass model. 

Our work shows that when the fraction of spectroscopic redshifts is small, and the lens inversion method is free to adjust source redshifts, a hitherto unexplored type of lensing degeneracies may become important, one that links the line of sight (sub)structure with source redshifts, in a way similar to the well known mass sheet degeneracy \citep{fal85}, which links the mass density in the lens with the density profile slope. In other words, a single lens plane mass distribution with model adjusted source redshifts can reproduce images as well as cluster mass distribution and line of sight structure, in combination with correct source redshifts. Therefore, lensing degeneracies can conceal line of sight structure.

How well the positions of observed images are reproduced by a model is commonly quantified by lens plane rms. Because lens model optimization tries to reduce lens plane rms, it is important how the rms is calculated.
Not all papers calculate the rms in exactly the same way; moreover, there are other possible estimators of the rms. As the lens mass models get progressively better, and the rms get smaller, it is important to adopt a common estimator, one that best reflects the differences between various mass models. In Section~\ref{calcrms} we consider several different formulations, and compare their values for the four mass maps generated in this paper. 

We adopt the concordance $\Lambda$CDM cosmological model: flat, matter density, $\Omega_m=0.3$, cosmological constant density, $\Omega_\Lambda=0.7$, and the dimensionless Hubble constant $h=0.7$. At the redshift of the cluster, $z_l=0.55$, 1\as~translates into 6.41 kpc. The critical surface mass density for lensing for sources ``at infinity'', $\Sigma_{\rm crit,\infty}=c^2/(4\pi G\cdot D_{\rm ol})=0.263$ g~cm$^{-2}$, where $D_{ol}$ is the angular distance from the observer to the lens.

\section{Previous work on MACS J0717}

MACS J0717 is one of the Cluster Lensing And Supernova survey with Hubble (CLASH; PI: M. Postman) clusters, and one of the 12 MACS clusters at $z>0.5$ \citep{ebe07}. It is extremely luminous in X-rays and is identified as a Sunyaev-Zel'dovich source, with one of its subclumps showing kinetic SZ effect \citep{say13}. \cite{ma09} found MACS J0717 to be an active triple merger with ICM temperatures exceeding 20 keV. The authors note that one of the three mergers is still on-going.

\cite{zit09} were the first to construct a lensing mass model of the cluster. They identified 13 multiply imaged systems, and build a parametric strong lensing mass model based on these. They found the cluster to have a shallow density profile and noted that this property will make it a very good cosmic telescope. As most papers, they use lens plane rms to quantify how well their model reproduces the positions on the sky of the observed images; they quote rms=2.2\as.

\cite{ebe04} discovered a large-scale, $\sim 4$~Mpc long filament connected to MACS J0717. The filament is seen as a pronounced overdensity of color-selected galaxies, well outside of the cluster's virial radius, which is $\sim 2.3$~Mpc. Spectroscopic follow-up showed that the filament is at the redshift of the cluster, and must be funneling matter to the cluster core.  \cite{jau12} used weak lensing analysis based on $10$\am$\times 20$\am~ACS mosaic to confirm the presence of a large-scale filament, at $3\sigma$.  \cite{med13} perform a joint strong and weak lensing analysis out to $5$~Mpc, and model the filamentary structure with 9 individual halos. They remark that while this is the most massive known cluster at $z>0.5$, its existence is not in tension with $\Lambda$CDM. More recent weak lensing analysis confirms the presence of large scale filament \citep{mar16}, and suggests despite the presence of numerous substructures, smooth accretion of surrounding material is the main source of mass growth in large clusters \citep{jau17}.

\cite{lim12} identified 15 multiply imaged systems in MACS J0717. Using the parametric modeling technique \lenstool, they constructed cluster mass model with 4 cluster-scale components. In fact, most later works agree that the cluster has 4 major components. The authors point out that the mass distribution in the cluster strongly deviates from that of the intra-cluster gas as traced by the X-ray surface brightness. Complex structure argues in favour of multiple mergers and ongoing dynamical activity, something that was already pointed out by earlier works. Their lens plane rms is about 2.5\as.

In 2013, MACS J0717 became part of the HFF project, and consequently the subject of more attention. It became the first target of the Grism Lens-Amplified Survey from Space (GLASS) survey; \cite{sch14} obtained and confirmed several high redshift multiply imaged source candidates behind the cluster. More recently, MACS J0717 mass reconstructions were performed by \cite{ric14} and \cite{joh14}, as part of an effort to model all 6 HFF clusters.  The latter authors used 14 image systems to obtain lens plane rms=0.38\as. \cite{die15} used a reconstruction technique that combines the mass from cluster galaxies with a flexible free-form description of mass on larger, cluster scale. They identified many new multiple images, and showed that the central density profile of this cluster is not well constrained due to the lack of images in that region. Their lens plane rms is 2.8\as.

\cite{zit15} analyzed MACS J0717 as part of a set of 25 clusters combining two different parametrizations of the strong lensing portion of the cluster with weak lensing constraints found at larger distances from the center. MACS J0717 stands out in their study. Of the 25 clusters, its model has by far the largest $\chi^2$/DOF, 2.70, and the largest lens plane rms, 3.18\as. The cluster also has the largest area inside Einstein radius, making it a very powerful, if difficult to characterize, cosmic telescope. 

\cite{lim16} showed that using either cored or peaky mass components to model the cluster-wide mass distribution provides equally good fits to MACS J0717, highlighting the fact that even HFF is not able to break lens modeling degeneracies in some locations within the cluster. The lens plane rms for their models ranges from 1.9\as to 2.4\as.

The model of \cite{kaw16} (K16), based on \glafic, uses a large number of strong lensing constraints---173 images from 60 sources---and produces a lens plane rms of 0.52\as. This low value might be attributable to the complexity of their model, which, in addition to mass due to many member galaxies, consists of 9 cluster-wide components, external shear, and 3 multipole perturbations, with the highest order one having the potential of the form, $\phi\propto r^2\cos[5(\theta-\theta_\star)]$. Their model also includes the mass of a bright foreground elliptical in the lower portion of the cluster, but places it at the redshift of the cluster. Because their model uses a large number of images and produces a low rms, we will use the K16 model as part of our analysis in this paper.

\section{Reconstruction method: \grale}

\grale~ is a free-form, adaptive mesh lens inversion method that does not rely at all on the light of the cluster member galaxies. It parametrizes the mass distribution with many projected Plummer profiles, whose width and mass are determined by an iterative procedure using a genetic algorithm. The distribution of the Plummer spheres across the face of the cluster is guided solely by the lensed images.  Mass maps that we present here are an average of 40 independent \grale~ runs, each started with its own random seed. Each run has of the order of a 1000-3000 Plummers, so the average has tens of thousands Plummers. The only inputs to \grale~are the locations and redshifts of lensed images.  \grale~ has been described extensively in previous works \citep{lie06,lie07,moh14}.

When reconstructing a cluster like MACS J0717, with potentially complicated mass distribution, it is important to note that free-form methods like \grale, that do not use any assumptions about light tracing mass (LTM), are immune to complications in the cluster mass distribution that might affect parametric models, like variations in the mass-to-light ratio of galaxies, or the appropriate number of cluster-wide mass components to include in one's model.

However, some factors present more of a challenge to free-form, non-LTM methods than to parametric methods. The foremost among these are missing source redshifts. \grale~ cannot easily constrain source redshifts, while parametric methods can. Having $20-50$ free parameters makes parametric models less flexible in terms of distributing mass, but endows them with predictive power. In \lenstool~ \citep{jul07}, for example, unknown redshifts can be included as additional parameters to be determined. Photometric redshift estimates are often used as priors, with typical uncertainties of  $\Delta z\approx \pm 0.5$. Extending the parameter set of \grale, with its $\mathcal{O}(10^3)$ free parameters, does not yield good constraints on source redshifts. To make \grale~predict source redshifts, one would need to run separate mass reconstructions with a range of trial redshifts, and then identify runs that produce better fitness measures, or better fits to the observed images. This procedure is cumbersome and time consuming. Therefore every source used in \grale~ reconstruction has to have a redshift as part of the input.

\section{Single lens plane reconstructions of MACS J0717}\label{singleLP}

In this section we will assume that all the lensing mass in the direction of MACS J0717 is at the redshift of the cluster, at $z_l=0.55$, so there is no need to consider multiple lens planes.

\subsection{\grale~ reconstruction using sources at all redshifts: \grale-all-$z$}\label{graleallz}

We used the data contributed by the HFF community (all of which has been published) as a starting point for source redshifts. Only 10 sources in MACS J0717 have spectroscopic redshifts. For the bulk of the sources whose spec-$z$'s are unknown, our adopted redshifts were based on the following: photo-$z$ from K16 and \cite{lim16}, model $z$'s from K16, and model $z$ from 3 models of \cite{lim16}. In all, up to 6 redshift estimates were available for some sources (i.e., image systems). For a given source these can disagree by as much as $|\Delta z|\sim 2$. The redshift estimates just from the 3 models of L16 often differ by as much as $0.6-1.2$. This is an indication that lens models adjust source $z$ values to fit the model priors, instead of predicting accurate $z$'s. We will come back to this issue in Section~\ref{lossec}. 

In our reconstructions we used sources that had at least 3 of the above mentioned 6 redshift estimates. For these, we calculated the dispersion in $z$, as $\delta z=2(z_{\rm{max}}-z_{\rm{min}})/(z_{\rm{min}}+z_{\rm{max}})$, where $z_{\rm{min}}$ and $z_{\rm{max}}$ were the smallest and largest of the redshift estimates. Out of these we selected sources with $\delta z\leq 0.4$, i.e. those with approximately consistent redshift estimates. This cutoff is arbitrary, but we settled on this value as it gave us a reasonably large number of images to base our reconstruction on. If a source satisfied this criterion, its $z$ was set to the average of the available $z$ estimates. This yielded 44 sources with 126 images. The redshift distribution of the sources is shown in the top panel of Fig.~\ref{zdistrib}.

The \grale~reconstruction\footnote{The input images for \grale~can be either extended, or point-like. In this case we used the extended image option, while for the reconstruction in Section~\ref{K16sec}, we used the point image option. When the number of images is large, as it is in these two cases, the difference between the two are small.} using these 126 images is called \grale-all-$z$. In Section~\ref{calcrms} we describe several different ways to calculate the lens plane rms, including the one we think is the most conservative estimator, and has the most discriminating power between mass models, eq.~\ref{rmstot}. However, the estimator most commonly used in the literature is eq.~\ref{rmsimsII}; its value for our \grale-all-$z$ map is 1.28\as.

Though we restricted ourselves to sources whose photo-$z$ and model predicted redshifts were not too discrepant, and took an average of the available redshifts, some of our assumed source redshifts probably deviated from the true ones by amounts that would affect the reconstruction. This likely contributes to the large rms (see Table~\ref{table1}). Because the source redshifts in this reconstruction were not adjusted by a lens inversion model (unlike the reconstructions in Section~\ref{K16sec}), and because we use sources at all the redshifts (unlike the reconstructions in Section~\ref{lossec}), we consider this reconstruction unbiased. This model also constitutes a compromise between the largest possible number of images, and the quality of the input data. 

\begin{figure*}    
\centering
\vspace{-5pt}
\includegraphics[width=1.0\linewidth]{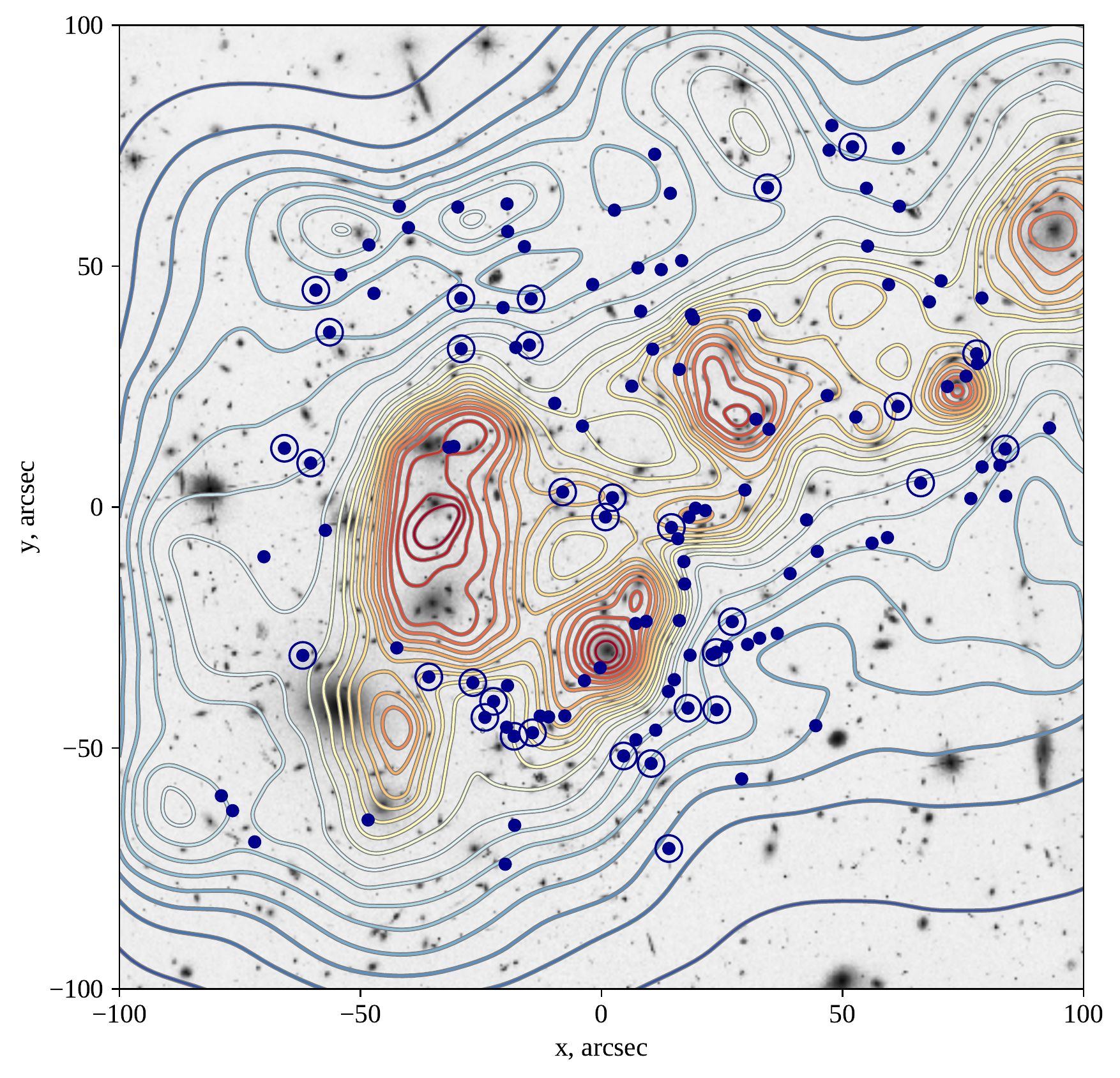}
\vspace{+5pt}
\caption{Contours of projected mass distribution of \grale-all-$z$ reconstruction of MACS J0717. The blue filled dots are the images used in the modeling. Circled dots represent sources with spectroscopic redshifts. The background HST Frontier Fields image is included to show the observed galaxies (credit: Judy Schmidt). This is the same reconstruction as that in the left panel of Fig.~\ref{massmaps1}.}
\label{massmaps0}
\end{figure*}

\begin{figure*}    
\centering
\vspace{-5pt}
\includegraphics[width=0.49\linewidth]{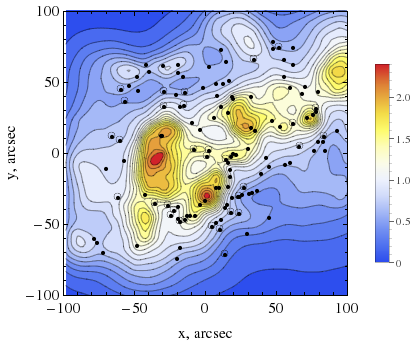}
\includegraphics[width=0.49\linewidth]{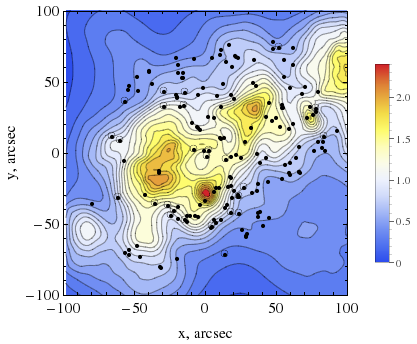}
\vspace{+5pt}
\caption{Contours of projected mass distribution of \grale~reconstructions of MACS J0717. The black filled dots are the images used in the modeling. Circled dots represent sources with spectroscopic redshifts. {\it Left:} The selection of input sources is described in Section~\ref{graleallz}, resulting in 126 images from 44 sources. {\it Right:} all K16 sources (60 sources and 173 images) were used in this reconstruction. The corresponding redshift distributions of the sources are shown in the top two panels of Fig.~\ref{zdistrib}. The color bar indicates projected density, in units of lensing critical surface mass density for sources ``at infinity'', $\Sigma_{\rm crit,\infty}$.}
\label{massmaps1}
\end{figure*}

The mass map is shown in Fig.~\ref{massmaps0}, with the HST Frontier Field observed galaxy field as the background, and in the left panel of Fig.~\ref{massmaps1}. The overall features of the map are consistent with those of other authors, for example, there are 4 main clumps, which are very similar to clumps A, B, C and D identified by \cite{lim16}.

\begin{figure*}    
\centering
\vspace{-5pt}
\includegraphics[width=0.49\linewidth]{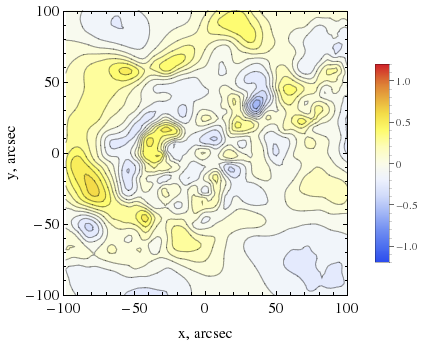}
\includegraphics[width=0.49\linewidth]{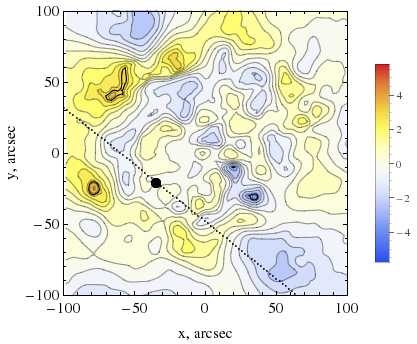}
\vspace{+5pt}
\caption{{\it Left:} \grale-K16 map (60 sources and 173 images; Section~\ref{K16sec}) subtracted from the \grale-all-$z$ map (44 sources and 126 images; Section~\ref{graleallz}). {\it Right:} map of the statistical significance of this difference map, with regions above $3\sigma$ outlined with solid thick curves. The large black dot is the center of the K16 map, and the dashed diagonal line is the orientation of the K16 external shear.}
\label{diffmaps1}
\end{figure*} 

\subsection{\grale~reconstruction using K16 image and redshift data: \grale-K16}\label{K16sec}

Since the spectroscopic redshifts for the bulk of the sources in MACS J0717 are not known, we would like to asses how uncertain redshifts affect the mass maps produced by \grale. To that end, we carry out another reconstruction of the cluster, but this time using the image positions used by K16 in their work, and source redshifts predicted by their lens model (see their Table 8). There are a total of 60 sources and 173 images.

Our goal in this subsection is not a comparison with K16 results, but that between the \grale~reconstructions of the two data sets of the same cluster, namely, the \grale-all-$z$, presented Section~\ref{graleallz}, and the one presented in this section, based on K16 data, which we call \grale-K16.\footnote{Note that K16 used $z_l=0.545$ in their work, while we use $z_l=0.55$. This difference will not affect the comparison results.} Using the same modeling software in both reconstructions eliminates it as a source of differences between maps. There are 97 images from 32 sources whose positions are in common with the 126 images of the \grale-all-$z$ map, and 173 images from K16. Among these 32 sources there are some differences in redshift. (Note that K16 redshifts were part of the input to estimate redshifts in Section~\ref{graleallz}.) The resulting mass map is shown in the right panel of Fig.~\ref{massmaps1}; its lens plane rms, as calculated using eq.~\ref{rmsimsII}, is 0.62\as. 

The two maps in Figure~\ref{massmaps1} have many features in common; the differences in the normalized surface mass density, $\Delta\kappa(\vec\theta)=\kappa_2(\vec\theta)-\kappa_1(\vec\theta)$, shown in the left panel of Figure~\ref{diffmaps1} are not very large. Here, subscript $1$ refers to the \grale-K16 map, while $2$ refers to the \grale-all-$z$ map. The right panel shows the contours of the significance of the difference, which we calculate as $\sigma(\vec\theta)=\Delta\kappa(\vec\theta)/\sqrt{[{\epsilon_1}^2(\vec\theta)+{\epsilon_2}^2(\vec\theta)]}$, where $\epsilon(\vec\theta)$ is the location dependent rms scatter between the 40 individual \grale~reconstructions. The regions above $3\sigma$ are delineated with thick lines; we will return to this difference in the maps in Section~\ref{hintlos}. In the region defined by the images the differences in the projected surface mass density are around $\pm 1\sigma$, and regions of positive and negative $\sigma$  are randomly distributed within the cluster. In other words, the differences in inputs of the two reconstructions---image positions and redshifts---do not lead to significant differences between the \grale-all-$z$ and \grale-K16 maps, especially in the image region.

We conclude that while \grale~is not very good at predicting redshifts or positions of additional images (not included in the reconstruction), its uncertainties account for possible extensions to, or small changes in the data set. Stated differently, \grale~maps with accompanying uncertainties are robust against reasonable changes in input data. This result echos that from \cite{pri17}, where \grale~uncertainties encompassed most other reconstructions of the two HFF clusters (their Figures 14 and 15).

\begin{figure}    
\centering
\vspace{-33pt}
\includegraphics[width=1.1\linewidth]{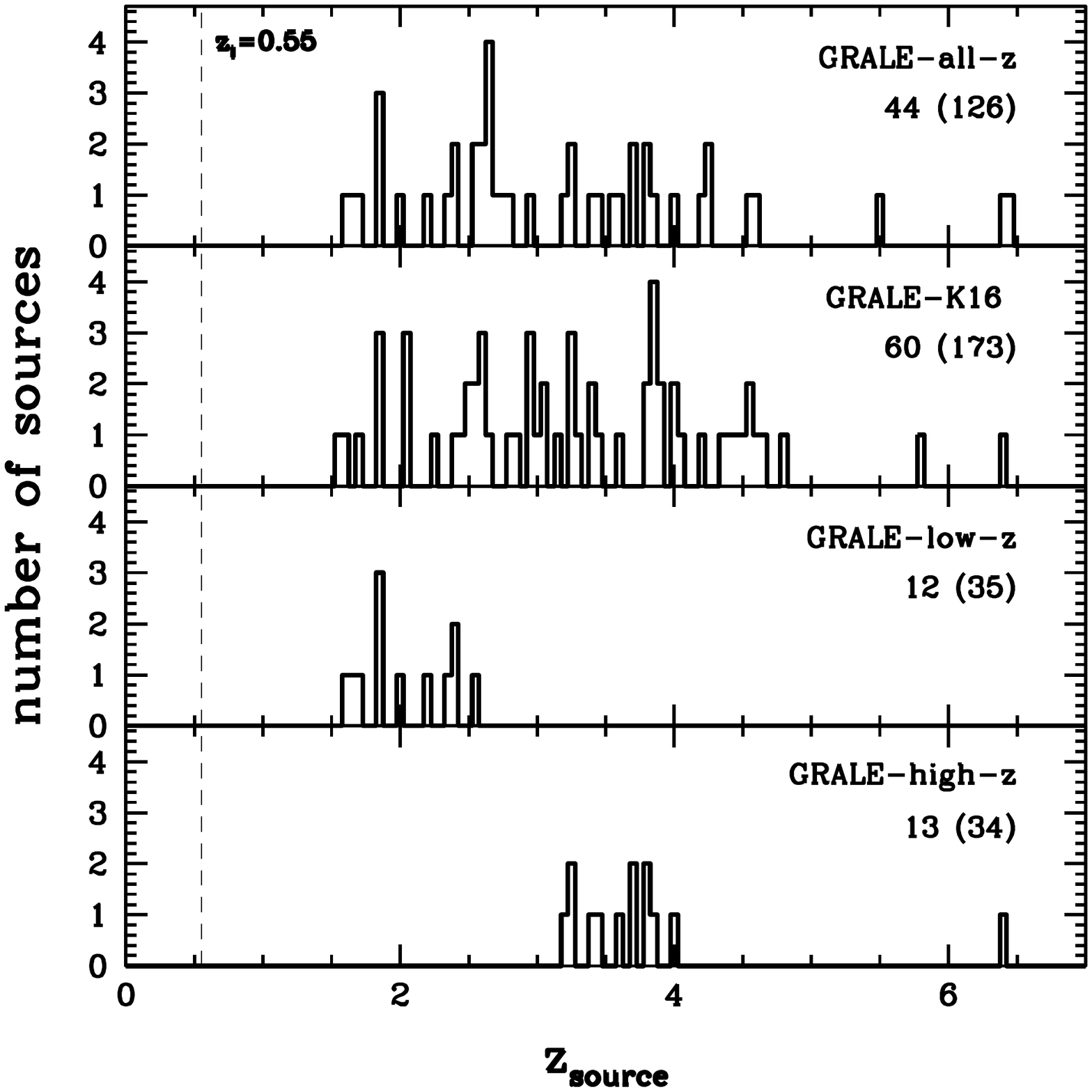}
\vspace{-75pt}
\caption{Redshift distribution of multiply imaged sources in the four \grale~reconstructions presented in this paper. The number of sources (images) is displayed in every panel. The cluster redshift is marked with a vertical dashed line. All \grale-K16 images, including the redshifts, are identical to the ones in the K16 model.}
\label{zdistrib}
\end{figure} 

\begin{figure*}    
\centering
\vspace{-5pt}
\includegraphics[width=0.49\linewidth]{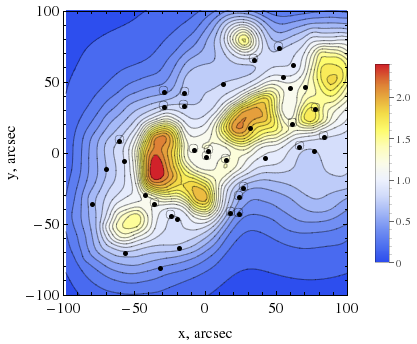}
\includegraphics[width=0.49\linewidth]{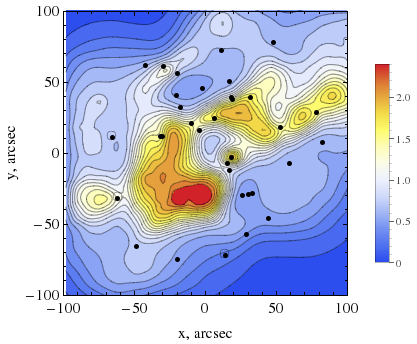}
\vspace{+5pt}
\caption{Contours of projected mass distribution of \grale~reconstructions of MACS J0717. The black filled dots are the images used in the modeling. Circled dots represent sources with spectroscopic redshifts. {\it Left:} Reconstruction using sources with redshifts $z_s<2.6$ (12 systems and 35 images). {\it Right:} Reconstruction using sources with redshifts $3<z_s<4.1$ (13 systems and 34 images), respectively. The corresponding redshift distributions of the sources are shown in the bottom two panels of Fig.~\ref{zdistrib}. The color bar indicates projected density, in units of lensing critical surface mass density for sources ``at infinity'', $\Sigma_{\rm crit,\infty}$.}
\label{massmaps2}
\end{figure*} 

\begin{figure*}    
\centering
\vspace{-5pt}
\includegraphics[width=0.49\linewidth]{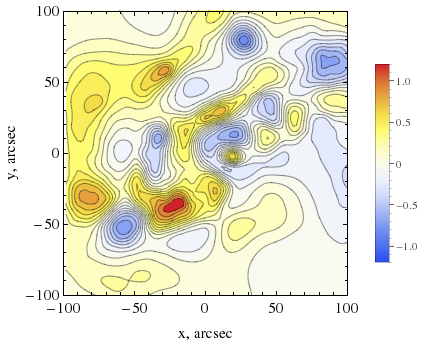}
\includegraphics[width=0.49\linewidth]{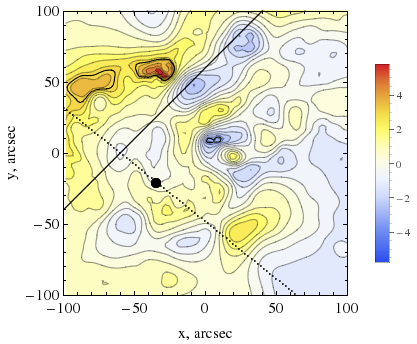}
\vspace{+5pt}
\caption{{\it Left:} The difference of the two mass maps presented in Fig.~\ref{massmaps2} (map that used lower redshift sources has been subtracted from the map that used high redshift sources). {\it Right:} The significance contours of this difference map, with regions above $3\sigma$ outlined with solid thick curve. The large black dot is the center of the K16 map, and the dashed diagonal line is the orientation of the K16 external shear. The region above and to the left of the solid straight line, $\Delta {\rm Dec}=\Delta {\rm RA}+60$\as, is used in Section~\ref{lossec} to calculate the significance of the difference between the two maps. The largest statistically significant region has positive mass excess in the map that uses $3.0<z_s<4.1$ sources. This suggests the presence of excess mass at roughly that location, and around $z\sim 2.5-3$. The color scale in both panels is the same as in Figure~\ref{diffmaps1}.}
\label{diffmaps2}
\end{figure*} 

\subsection{Hint of LoS?}\label{hintlos}

The largest and most significant differences between the two maps in Fig.~\ref{massmaps1} are in the upper left of the map (see Fig.~\ref{diffmaps1}); \grale-all-$z$ shows a mass excess compared to \grale-K16. In Sections~\ref{lossec} and \ref{examK16} we will argue that it is due to a line of sight structure, located significantly behind the cluster. 

Because the location of the mass excess is somewhat outside of the image region, only the polar direction with respect to the cluster center, in the plane of the sky, towards the mass excess is well constrained by the reconstruction. Its precise distance from the cluster center on the plane of the sky, as well as its shape, are not well constrained. The total mass in the two maps differs by $\sim 3\times 10^{13}M_\odot$, with the \grale-all-$z$ having the larger mass. (The actual mass will depend on the sky-projected distance from the cluster.) In Section~\ref{examK16} we suggest that in the K16 model, this excess mass, which is largely external to the main cluster, is partly accounted for by external shear. 

There are other hints as well. It has been noted by some authors that MACS J0717 is a difficult cluster to model. This is reflected in relatively large lens plane rms for some models. Furthermore, the model-predicted redshifts from the literature tend to differ considerably, even within a single family of models, and/or reconstruction technique, and some model-predicted redshifts differ from photometric ones. It is possible that these modeling difficulties and discrepant predicted redshifts are the result of models---not modelers!---attempting to account for the line of sight structure that is not explicitly present in the models.

The line of sight towards MACS J0717 has a few peaks in the redshift distribution of sources: Figure 3 of  \cite{med13} shows the BPZ-based redshift distribution up to z=4 of $z^\prime<25$ galaxies. There are a few possible peaks, at $z$ of about 0.9, 1.4, 2.1, and 2.8.

These are all reasons to investigate further. In the next section we test our hypothesis of LoS structure using additional \grale~reconstructions, and in Section~\ref{examK16} we show that the K16 reconstruction suggests the presence of LoS. 

\section{Reconstructions using image sets disjoint in redshift: \grale-\lowercase{low-z} and \grale-\lowercase{high-z}}\label{lossec}

One way to handle clusters with LoS structures using a single lens plane formalism is to carry out separate reconstructions using sources in different, non-overlapping redshift ranges. If a significant LoS structure is present, a mass reconstruction using sources with redshifts $z_1\rightarrow z_2$ will see less mass than that using $z_3\rightarrow z_4$, if $z_3>z_2$. Subtracting the former mass map from the latter will reveal structure at intermediate redshifts. While this is not a fully correct method of treating LoS because all mass is assumed to be at a single $z_l$, it can identify the presence of LoS structure behind the main cluster. This method was used in \cite{moh14} for Abell 1689.

The redshift limits of our two redshift ranges in the case of MACS J0717 were picked as follows. We excluded all sources above  $z=4.1$ because the number of sources at these high redshifts is small, and the probability of line of sight structure increases with redshift. We also noticed that the largest number of sources per redshift range is highest around $z\sim 2.7$, similar to the redshift of one of the peaks in \cite{med13}, which may indicate the presence of excess mass at these redshifts. Obviously, these are not very strong arguments, so our choice of source redshift ranges is somewhat arbitrary.  For the purposes of a fair comparison of the two reconstructions, we would like to have roughly equal number of sources and images in each of the two redshift ranges. Guided by this, we chose the lower source redshift range to have $z_1=0$ and $z_2=2.6$, giving us 12 systems with 35 images, and the higher redshift range to have $z_3=3$ and $z_4<4.1$, containing 13 systems with 34 images. The two reconstructions are called \grale-low-$z$ and \grale-high-$z$, and the corresponding surface mass densities are $\kappa_{\rm lo}(\vec\theta)$ and $\kappa_{\rm hi}(\vec\theta)$. (In the $3<z_s<4.1$ range we did not use source 74, because removing it improved \grale~fitness considerably. Its two photo-$z$'s are 4.5 and 3.8, and the average redshift, calculated using the procedure described above gives a much lower value of 3.54.) The gap between the two redshift ranges omits 9 sources (28 images). 

The reconstructions using these 2 different source redshift ranges are shown in Figure~\ref{massmaps2}: the left (right) panel presents the \grale-low-$z$ (\grale-high-$z$) maps. The difference between the two reconstructions, $\Delta\kappa(\vec\theta)=\kappa_{\rm hi}(\vec\theta)-\kappa_{\rm lo}(\vec\theta)$, is shown in the left panel of Fig.~\ref{diffmaps2}. (The definition of these quantities are exactly the same as those in Section~\ref{K16sec}.) Yellow/red (blue) colors represent regions of density excess (deficit). If the reconstructions were perfect, this map should not have any negative (blue) contours at all because higher redshift sources should ``see'' all the mass that lower redshift sources ``see'', plus possibly more. To interpret this map one needs to identify regions where the density difference is statistically significant. The right panel shows the contours of the significance of the difference, which we calculate as $\sigma(\vec\theta)=\Delta\kappa(\vec\theta)/\sqrt{[{\epsilon_{\rm hi}}(\vec\theta)^2+{\epsilon_{\rm lo}}(\vec\theta)^2]}$, where $\epsilon(\vec\theta)$ is the rms scatter between the 40 individual \grale~reconstructions of the corresponding map. The regions above $3\sigma$ are delineated with thick lines. The most significant regions are in the upper left of the panel, and the positive density excess region is the most prominent. We interpret this to mean that there is likely excess mass, probably to the side of the line of sight to the cluster. The mass excess is outside the main image region, and roughly in the same location as that seen in Section~\ref{hintlos}. As we noted in that Section, the exact location, i.e., sky-projected distance from the cluster center, and shape of the mass excess are not well constrained; however, its existence and direction with respect to the cluster center are. It's redshift is around $2.5-3$. 

The total masses within the square region shown in Fig.~\ref{massmaps2} are $1.46\pm 0.025\times 10^{15} M_\odot$ and $1.58\pm 0.041\times 10^{15} M_\odot$, for the \grale-low-$z$ map and \grale-high-$z$ maps, respectively, hence high redshift sources detect about $1.2\times 10^{14} M_\odot$ more mass than the low redshift sources. The robustness of this mass differences can be assessed by estimating it in a different way. Considering only the areas (of positive and negative mass excess) where $|\sigma|>2$ and $|\sigma|>3$, yields smaller mass differences of $4.18\times 10^{13} M_\odot$ and $1.79\times 10^{13} M_\odot$, respectively. To determine where, on the sky, most of the mass difference resides, we restrict this calculation to the portion of the lens plane above the line $\Delta {\rm Dec}=\Delta {\rm RA}+60$\as~(shown as the solid black line in the right panel of Fig.~\ref{diffmaps2}). The two resulting mass differences are $4.85\times 10^{13} M_\odot$ and $1.88\times 10^{13} M_\odot$, showing that most of the mass excess is to the upper left of the main cluster. These two values are larger than the two quoted above because the area above the $\Delta {\rm Dec}=\Delta {\rm RA}+60$\as~line has mostly regions with $\sigma>0$, rather than $\sigma<0$. All these estimates suggest that there is at least $10^{13} M_\odot$ more mass detected by the \grale-high-$z$ map, and the bulk of that mass is located in the upper left corner.  

\begin{figure*} 
\centering
\vspace{-70pt}
\includegraphics[width=0.99\linewidth]{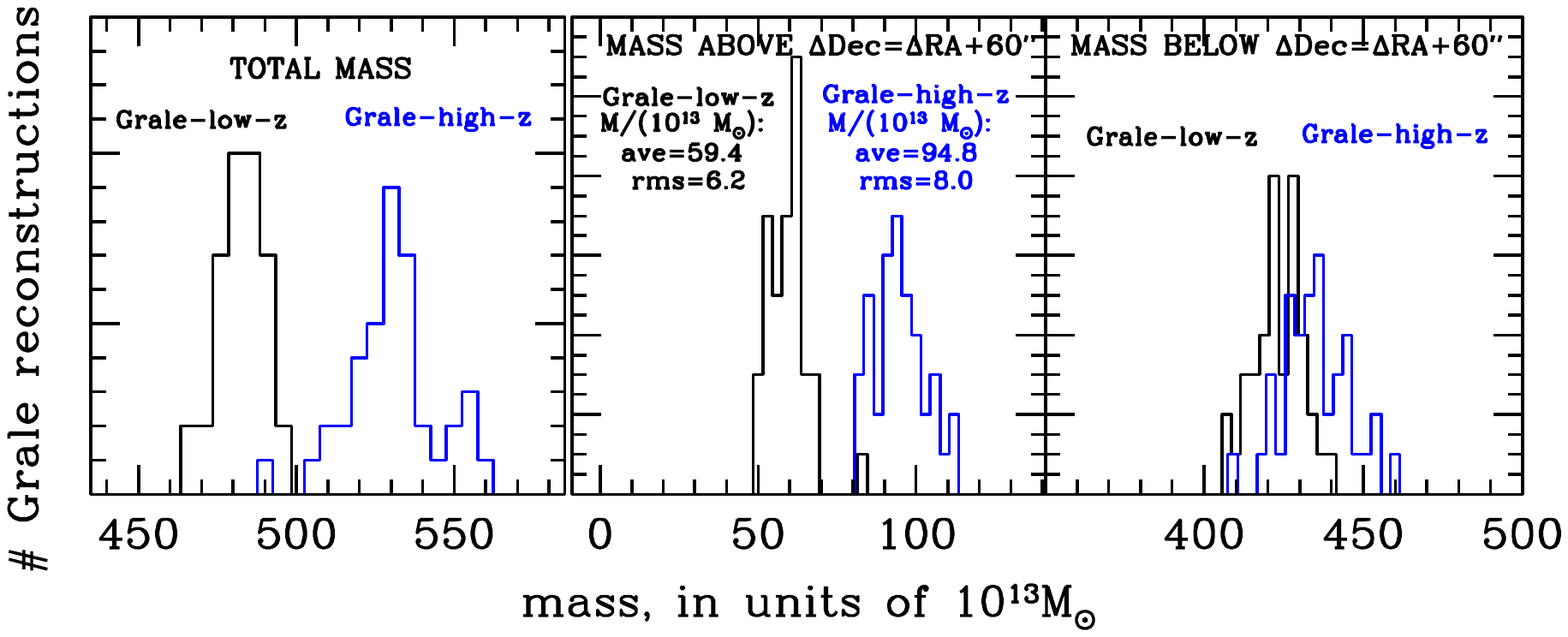}
\vspace{-370pt}
\caption{{\it Left:} The total mass, in units of $10^{13} M_\odot$, of 40 \grale-high-$z$ (blue) and 40 Grale-low-$z$ (black) independent reconstructions. {\it Middle:} The mass in the 40 \grale-high-$z$ (blue) and 40 Grale-low-$z$ (black) reconstructions above the $\Delta {\rm Dec}=\Delta {\rm RA}+60$\as line. The values of the average and rms of the masses are shown in the panel. {\it Right:} The mass in the 40 \grale-high-$z$ (blue) and 40 Grale-low-$z$ (black) reconstructions below the $\Delta {\rm Dec}=\Delta {\rm RA}+60$\as line. The comparison of the three panels shows that the mass excess is in the \grale-high-$z$ maps, and it is confined to the region above the $\Delta {\rm Dec}=\Delta {\rm RA}+60$\as line.}
\label{fortymaps}
\end{figure*}

To ensure that the mass excess in the upper left of the \grale-high-$z$ map, compared to the same region in the \grale-low-$z$ map, is not due to just a few of the \grale~reconstructions, we calculate the mass in each of the 40 independent \grale-low-$z$ and \grale-high-$z$ reconstructions, to the upper left of the $\Delta {\rm Dec}=\Delta {\rm RA}+60$\as~line, and compare these to the mass in the rest of the map. The histograms of these values are shown in the middle and right panels of Figure~\ref{fortymaps}. The left panel shows the total mass in each of the 40 \grale-low-$z$ (black) and 40 \grale-high-$z$ (blue) maps. Comparison of all the histograms shows that the mass excess exists in 39 out of 40 maps, and is confined to the upper left corner of the \grale-high-$z$ maps.

The K16 model includes a large amplitude external shear, $\gamma_{\rm K16}=0.12$, whose axis is represented by the straight dotted line in the right panel of Fig.~\ref{diffmaps2}. The large filled dot on that line is the center of their cluster. The orientation of the shear axis is consistent with the direction (on the plane of the sky) towards our excess mass. (Note that \grale~does not have external shear.)  A $5\times 10^{13} M_\odot$ point mass would produce $\gamma=0.083$ at a distance of $50$\as, for sources at $z_s=3$.  In other words, the external shear of the K16 model and the external mass we propose have similar orientation and shear magnitude; the difference is that for the point mass $\gamma$ falls as $|\theta|^{-2}$, while external shear has the same magnitude at all separations. Though our excess mass is likely at $z\sim 2.5-3$, while K16 model shear is at cluster redshift, we suspect that their shear helps to account for the same mass that we find.

\begin{figure*}    
\centering
\vspace{-70pt}
\includegraphics[width=0.99\linewidth]{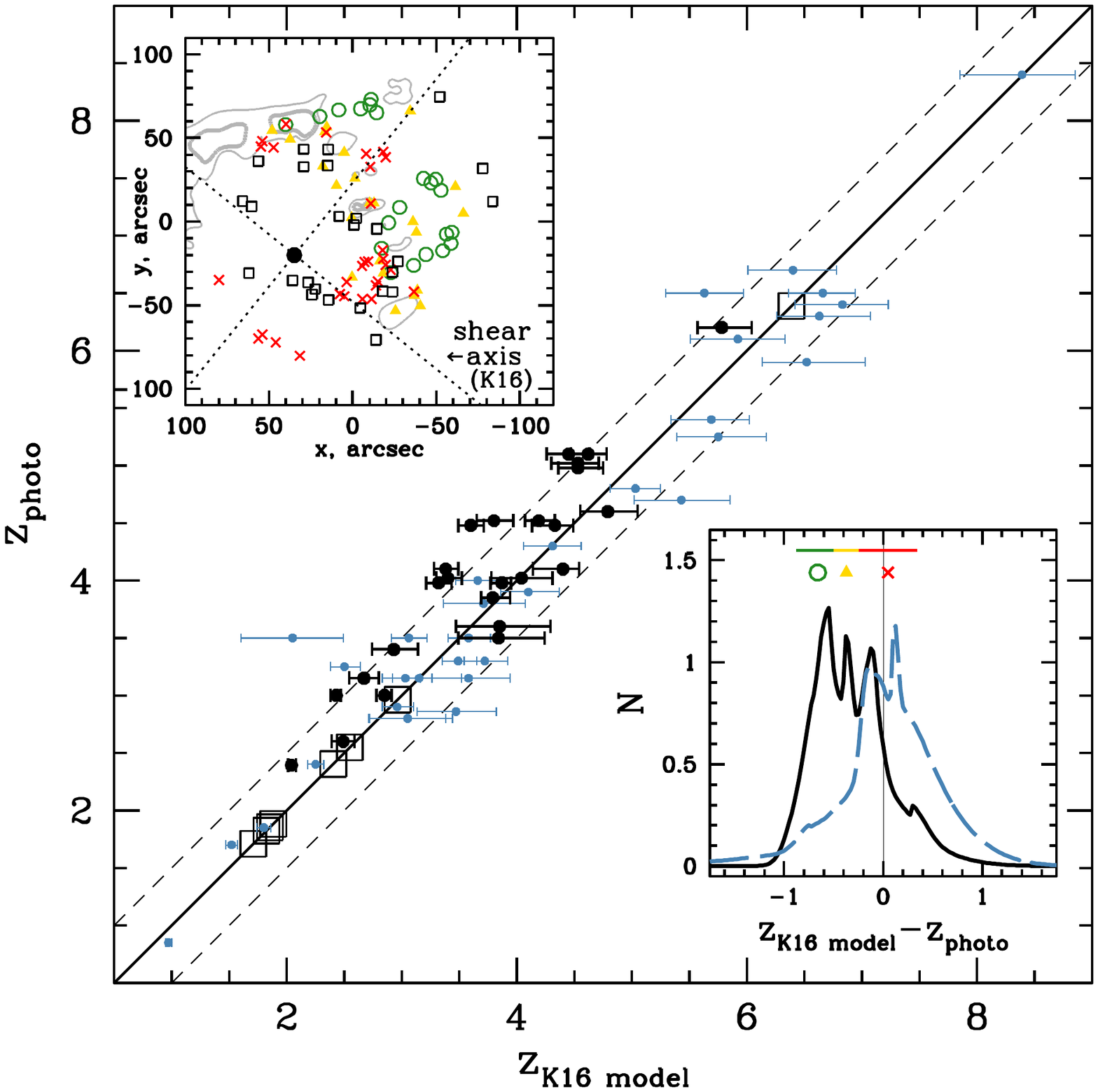}
\vspace{-115pt}
\caption{Main plot: Black filled dots represent MACS J0717 sources with photometric redshifts, whose K16 model predicted redshifts are also shown. Horizontal error-bars are model uncertainties. The dashed diagonal lines represent photo-$z$ uncertainties.. Black empty squares are the 8 sources with spectroscopic redshifts. Since the publication of K16, one additional system, \#6, had its spec-$z$ measured. We plot its  spec-$z$ on the vertical axis, instead of its photo-$z$; its location in this plot is (2.04, 2.393). Light blue points show photo-$z$ and K16 model fitted redshifts for sources behind Abell 2744, MACS J01416, and MACS J1149. The lower right inset summarized the data from the main plot in the form of two histograms: black for MACS J0717, and light blue dashed for the other 3 clusters modeled by K16. The upper left inset shows the sky distribution of spec-$z$ sources (black empty squares), and photo-$z$ sources (color and shape indicates the value of $z_{\rm{model}}-z_{\rm{photo}}$; see the bars and corresponding symbols above the black solid histogram in the lower right inset). The gray contours are significance levels of $\sigma=2$ and $3$, from the right panel of Fig.~\ref{diffmaps2}.}
\label{K16photoz}
\end{figure*} 

\section{Examination of the K16 model results}\label{examK16}

Further evidence supporting LoS structure at intermediate redshifts comes from the K16 model.

The main portion of Figure~\ref{K16photoz} shows, as large black dots with error-bars, 24 sources that have photo-$z$, plus one source whose spectroscopic redshift was determined after the K16 work was published. These are plotted against redshifts predicted by the K16 model, using photo-$z$ priors. (We took the liberty to vertically displace some of the points by $\Delta z=\pm 0.02$ to avoid overcrowding.) The horizontal error bars show lens model uncertainties. The diagonal solid line is the 1-to-1 correspondence, while the two diagonal dashed lines show the uncertainties in the photometric redshifts, $\pm 0.5$. The large empty black squares with no error-bars represent the 8 sources with spectroscopic redshifts. (Three sources have $z=1.855$; we displaced these a little to separate them out.)

The most notable feature is that K16 model consistently underpredicts source $z$'s compared to their photometric estimates for MACS J0717.  Contrast that with the K16 model $z$ vs. photo-$z$ for the sources behind the other 3 clusters K16 analyzed: Abell 2744, MACS J01416, and MACS J1149, shown as the light blue points with error-bars. These show no systematic displacement from the diagonal line. 

A summary of these data is presented as a histogram in the lower right inset. The horizontal axis is the difference between the model and photo-$z$. Each model redshift has been smeared into an asymmetric Gaussian, whose left and right half-widths are the corresponding K16 model uncertainties. MACS J0717 sources make up the solid black histogram, while the sources from the other 3 clusters are shown as the light blue dashed histogram. Though not shown individually to avoid overcrowding, each of the 3 cluster histograms, for Abell 2744, MACS J01416, and MACS J1149, peak very close to zero. While this latter distribution (light blue) is centered on zero, the former (black) is displaced, indicating that model redshifts for sources behind MACS J0717 are consistently lower than corresponding photo-$z$'s. 

A further peculiar property of these sources is illustrated in the upper left inset of Figure~\ref{K16photoz}. The green empty circles, yellow filled triangles and red crosses show images whose $z_{\rm{model}}-z_{\rm{photo}}$  are $<-0.5$, between $-0.5$ and $-0.25$, and $>-0.25$, respectively (see corresponding horizontal bars over the black solid histogram in the lower right inset). Their distribution on the sky is not random: sources with most discrepant model $z$ and photo-$z$ are furthest from the cluster center, represented by a large filled dot. Those with intermediate and smallest discrepancies are closer to the center. The gray contours represent the significance levels of the $\Delta\kappa$ mass distribution, from the right panel of Fig.~\ref{diffmaps2}, shown here to orient the eye. The two dotted lines are the external shear axes of K16.

We suggest that the consistent underestimation of source redshifts by the K16 model, as well as their non-random spatial arrangement described above compensate for a systematic difference between the actual mass distribution in the cluster and along the line of sight, versus the single lens plane mass distribution of the K16 model. 

We speculate that if all the lensing mass of MACS J0717, including that not directly along the light of sight, were at the redshift of the cluster, then a sophisticated and quite flexible lensing inversion method such as \glafic~would be able to reproduce it without systematic changes to source redshifts. After all, this is the case for each of the 3 other clusters: Abell 2744, MACS J01416, and MACS J1149. Modeling becomes more problematic when there is LoS structure that affects sources with different $z_s$ differently.

Can we place constraints on the location of the LoS structure from the clues provided by the K16 model?  Probably yes. Recall that image separation is determined by the amount of mass in the lens, as well as source redshift. Large separations imply larger mass, or higher $z_s$. If the additional LoS mass were directly behind the main cluster, at $z_{\rm LoS}>z_l$, and had roughly the same shape when projected on to the sky, one would expect that to compensate for it within a single lens plane model at $z_l$, one would need to place sources that are behind the LoS structure at redshifts higher than their actual redshifts, because the additional converging mass at $z_{\rm LoS}$ is not present in the mass model. (No external shear would be needed to be included in this case.)

However the opposite is seen in Figure~\ref{K16photoz}; K16 model places some sources at lower than their photo-$z$'s, and this effect is most pronounced for images (green circles) that are far from the K16 origin. We propose a possible, approximate interpretation of how a single-plane model, like that in K16, can model a multi-plane lens like MACS J0717. Our interpretation does not explain all the features of K16, and many variations on the basic scenario proposed here are also possible.

Let us assume that there is indeed an LoS structure to the upper left of the cluster, at $z_{LoS}\sim 2.5-3$, as suggested by our \grale~modeling, and shown as gray contours in the upper left inset. K16 model does not explicitly include it. 

The action of this LoS mass, which is not along the same line of sight as the cluster can be approximated by external shear, such as the one in the K16 model, $\gamma_{\rm K16}$, at $z_l$. Because the sources foreground to the LoS mass do not see that additional mass, the action of $\gamma_{\rm K16}$ would need to be canceled for these sources, by something else within the model. 

First we address how this might work for the sources further away from the cluster center, and in the next paragraph, for the sources closer to the center. Close to where the LoS structure resides, the deflection angles due to $\gamma_{\rm K16}$ and those due to the cluster are approximately oppositely directed, so to cancel the effect of the shear, the cluster mass in that region has to be increased. This increase in mass is also seen by the more distant sources, background to the LoS. To reproduce their observed image separations, their redshifts have to be decreased. We speculate that this is the reason for the decreased source redshifts in the K16 model seen in Fig.~\ref{K16photoz}.

Closer to the cluster center the deflection angles from the $\gamma_{\rm K16}$ shear are smaller, and the cluster mass is larger, so there is no need to cancel the effect of $\gamma_{\rm K16}$ for $z<z_{\rm LoS}$ sources, and the cluster mass in this region would be close to its actual mass. Furthermore, there are sources in this region with spectroscopic redshifts (black empty squares in the main portion and the upper left inset of Fig.~\ref{K16photoz}), which ensure that the mass normalization of the model is correct. Since the projected mass due to the main cluster is large in this region, the relative influence of the external mass is smaller, even for sources at $z>z_{\rm LoS}$.

Though the scenario just outlined does not perfectly explain all the features of the model, we suggest that there exists a scenario that plays the source redshifts and the LoS structure off each other; in other words, the two sets of parameters are degenerate.

We note that systematic differences between measured and modeled redshifts have been noted by other authors. For example, \cite{cam16} show that if all redshifts are left free to be adjusted by a model, these tend to be higher than the observed ones, and the discrepancy grows with redshift. The authors do not comment why the model predicted redshifts are higher than the actual ones, but we note that this is in the opposite sense to what we see here. \cite{joh14} performed a somewhat different experiment, measuring redshift discrepancies for sources around images with observed or modeled redshift, in HFF cluster AS 1063. They also find systematic redshift differences in observed vs. model predicted redshifts.

\section{Discussion and Conclusions}\label{disc}

The influence of LoS mass on cluster strong lensing has been discussed in a number of papers. Based on the power spectrum of the large scale density fluctuations, \cite{hos12} shows that lensed images could have typical relative deflections of $1$\as$-2.5$\as, for clusters at intermediate redshifts. 
\cite{dal11} use the Millennium simulation halo catalogue \citep{spr05} for their LoS structures, and conclude that unaccounted for LoS halos can introduce deflections with respect to the cluster-only models, of as much as few arcseconds, for clusters at $z_l\sim 0.2-0.3$. \cite{chi17} analyze mock analogues of MACS J0416 placed along different lines of sight with 11 intervening halos, and estimate that LoS contributes $\sim 0.3$\as~to the lens plane rms.  \cite{cam16} used toy models to examine the effect of up to 10 mostly foreground massive galaxies on the lens plane rms, and conclude that $0.3$\as~ is the typical effect. Since the latter two studies use a very limited number of LoS halos, relative deflections of $1$\as$-2$\as~due to LoS seem more realistic. 

Given the large angular extent of MACS J0717, it seems unlikely that it has no intervening LoS; for example, \cite{med13} find several spikes in the redshift distribution of objects towards the cluster. As described in the introduction, the lens plane rms of the various MACS J0717 lens models span a wide range, from $\sim$0.4\as~to $\sim$3\as. It is possible that the LoS contributes to the rms of the models that have rms towards the upper end of that range. But how does one explain the small rms values, especially if the number of images used in the model is large (making the images more difficult to fit)? 

In this paper we concentrated on one model---that of \cite{kaw16}---because it used a large number of images (60 sources and 173 images), and obtained a very good fit to the image positions (rms $=0.52$\as). Our own \grale-K16 reconstruction based on K16 data yields a similarly low value of $0.62$\as. Since K16 presented detailed information on their model, we were able to examine it in detail. We suspect that our conclusions are applicable to other models that do not explicitly take LoS into account, and use their lens models to find redshifts for sources with photometric or unknown redshifts.

We present arguments suggesting that low rms are possible even in the presence of LoS structures, because the lens inversion method is able to adjust redshifts---typically by $\Delta z_s\!\sim\!\pm\,0.5$---of most sources to construct a model without multiple lens planes. In other words, if most source redshifts are not fixed by spectroscopy, the space of lensing degeneracies is large enough to include single lens plane solutions that reproduce the image positions as precisely as the actual mass distribution with multiple lens planes. These degeneracies are quite different from the well known mass sheet degeneracy, which is broken by sources at multiple redshifts. They are also different from other families of degeneracies, described in \cite{sah00}, \cite{lie08}, \cite{lie12}, \cite{ss14}, and \cite{pri17}. To our knowledge, the existing literature does not contain a theoretical discussion of degeneracies where source redshifts are allowed to vary.

If the fraction of spectroscopically determined redshifts is small, as it is in the case of MACS J0717, the mass solution found may be systematically different from the true map. Some of these differences are due to the unaccounted for LoS structure, and will lead to biases in the magnification estimates of high redshift sources, which are of paramount interest to the HFF project. Foreground structure, while potentially having more impact \citep{mcc17}, can be included in modeling because it is often bright enough to be detected. Background structure, especially at redshifts above $1-2$  is not easily detected, and poses more of a challenge. We suggest that one way of looking for it is to compare model estimated redshifts with photometric ones, as we do in this paper.   

In the case of MACS J0717 examined here, this analysis lead to a detection of a putative mass excess at redshifts around $2.5-3$. While the plane-of-the-sky distance to it is less certain (likely to be 75\as or larger), the position angle is better constrained, and coincides with the external shear axis of the \cite{kaw16} model. 

Our main conclusion is that allowing the lens model to adjust source redshifts can conceal LoS structure.This is a consequence of lensing degeneracies, many of which are not broken by the numerous multiple images at different redshifts. K16 reconstruction shows that if the fraction of known spectroscopic redshifts is small (in this case $\sim 20\%$), one can find redshifts for other sources that would be optimal for a single lens reconstruction, but these are not necessarily correct.

\section*{Acknowledgements}

The authors are grateful to all the Hubble Frontier Fields modeling teams for providing lensed image data. 
The authors acknowledge the Minnesota Supercomputing Institute for their computing time, resources, and support.
The authors would like to thank Judy Schmidt for providing the background HST image for Figure~\ref{massmaps0}.




\appendix

\section{Calculating lens plane rms}\label{calcrms}  

\begin{table*}
\centering
\begin{tabular}{|c|c|c|c|c|c|c|c|c}
        \hline {\bf Model}   &{\bf \# sources}&{\bf \# images}& $M(<\!{\rm crit},z_s\!=\!2.5)$&$\Delta^{\rm I}_{\rm rms,tot}$&$\Delta^{\rm I}_{\rm rms,src}$&$\Delta^{\rm I}_{\rm rms,ims}$ & $\Delta^{\rm II}_{\rm rms,ims}$ & $\Delta^{\rm II}_{\rm rms,avg}$\\
        \hline \grale-all-$z$   & 44         & 126       & $2.68\pm 0.14\times 10^{14}$ & 1.88 & 1.82 & 1.22 & 1.28 & 0.87\\
        \hline \grale-K16       & 60         & 173       & $2.42\pm 0.13\times 10^{14}$ & 0.94 & 0.83 & 0.57 & 0.62 & 0.42\\
        \hline \grale-low-$z$   & 12         &  35       & $3.01\pm 0.27\times 10^{14}$ & 0.30 & 0.30 & 0.20 & 0.20 & 0.18\\
        \hline \grale-high-$z$  & 13         &  34       & $4.02\pm 0.24\times 10^{14}$ & 0.69 & 0.68 & 0.41 & 0.53 & 0.41\\
        \hline \cite{kaw16} (K16)& 60        & 173       &     --                      & --   & --   & --   & 0.52 & -- \\
        \hline \cite{joh14}     & 14         &  42    & $5.91^{+0.2}_{-0.08}\times 10^{14}$& --   & --   & --   & 0.38 & -- \\
        \hline \cite{zit09}     & 13         &  34       & $7.4\pm 0.5\times 10^{14}$   & --   & --   & --   & 2.2  & -- \\
\end{tabular}
\caption{Summary of models. The first four lines are models calculated in this paper, while the last three are from the literature. We list the number of strongly lensed sources and corresponding images used to create the models. The mass, in solar masses, is that enclosed within the critical $\kappa=1$ contour for sources at $z_s=2.5$. The last five columns show the lens plane rms values calculated using estimators of eq.~\ref{rmstot}, \ref{rmssrc}, \ref{rmsims}, \ref{rmsimsII} and \ref{rmsavg}, respectively.}\label{table1}
\end{table*}

\begin{figure*}    
\centering
\vspace{-70pt}
\includegraphics[width=0.99\linewidth]{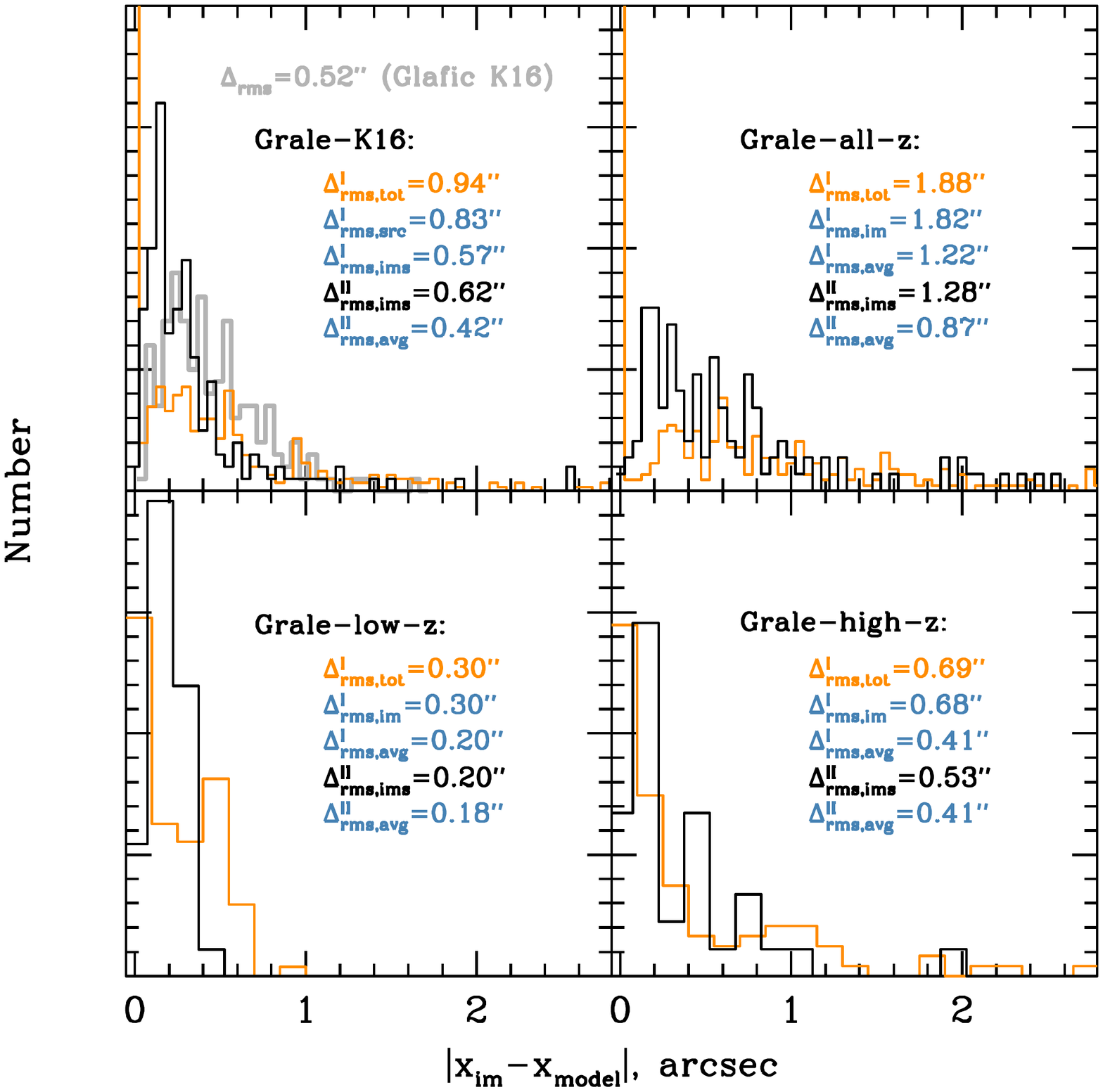}
\vspace{-115pt}
\caption{Lens plane rms. Each panel represents one of the four \grale~reconstructions presented in this paper. Since the number of images in these reconstructions is not always the same, the area under the histograms has been renormalized to the same in the two upper and the two lower panels. The upper left panel contains runs that were done using K16 data set. Because these can be directly compared to the \glafic~results of K16, the LP rms distribution from that paper (their Fig.4, lower left) is reproduced here as the gray thick histogram. The other two distributions (orange and black) in each panel represent data going into the calculation of eq.~\ref{rmstot} and \ref{rmsimsII}, described in Section~\ref{calcrms}. All 5 of the rms estimator values from that Section---eq.~\ref{rmstot}, \ref{rmssrc}, \ref{rmsims}, \ref{rmsimsII} and \ref{rmsavg}---are quoted in each panel. Some histograms have a few points extending beyond the right limit of the plot; these are not shown, but are included in the quoted rms. The black histogram and \ref{rmsimsII} are what is generally quoted in the literature. (Note: to evaluate eq.~\ref{rmstot}, each observed image has $J_i$ model predicted images, where $J_i$ is the image multiplicity of that source (typically $J_i=3$). One of these coincides with the observed image; this give rise to the spike in the orange histogram at $|\vec x_{\rm im}-\vec x_{\rm model}|\approx 0$. All model predicted images are included in the histogram, which is then normalized to have the same area.)}
\label{LPrms}
\end{figure*} 

Given a model surface mass density distribution, one can quantify how well it reproduces the observed image positions, by calculating the lens plane rms. (For a calculation of source plane rms see \cite{ogu10}.) While we argue in this paper that a small lens plane rms is a necessary but not sufficient condition for a model to be considered good, two questions remain: how small is small enough, and what definition of lens plane rms to use. 
Because rms in used in optimization to determine the best solution, and because recent reconstructions produce rather small rms values, the exact definition of rms is important. On could imagine that some definitions---like the ones we discuss in (I) below---could be more sensitive to substructure (in the cluster, or LoS), than others. A more detailed examination of how rms is affected by the lens mass distribution would require a separate paper.

It appears that not all papers in the literature are using the same definition of lens plane rms. In this section we write down a few possible definitions, and compare the values for the four \grale~reconstructions presented in this paper.

We also note that in contrast to most (if not all) other methods, \grale~does not minimize the image rms to find solutions, but optimizes a set of other quantities encapsulated in the fitness values. These quantify how well images from a given source overlap in the source plane, and how often spurious images are produced. The lens plane rms values are calculated only after a mass model has been produced. The rest of the discussion in this Section is applicable to any type of mass model.

Each reconstruction has $i=1,...I$ sources, and each source has $j=1,...J_i$ observed images. The location of each image is denoted by ${\vec\theta}_{i,j}$.  The total number of lensed images in the whole cluster is $J=\sum\limits_{i=1,I}J_i$. 

Each observed image $j$ of source $i$, ${\vec\theta}_{i,j}$ is lensed back to the source plane using the deflection angles calculated from the mass distribution. From here on, there are two different ways to proceed, (I) use each of these backprojected images as a source itself, or (II) average the positions of these backprojected images belonging to the same source, in the source plane, to obtain a single model-predicted source.

(I) Each of the $j=1,...J_i$ backprojected images in the source plane is used as a source itself, i.e. lensed forward, or relensed, to the lens plane, producing $J_i$ model predicted images per each observed image, and $K_i=J_i^2$ model predicted images per source. Their locations are designated by ${\vec\theta}_{i,j,k}$. One of the $J_i$ relensed images should coincide exactly with the corresponding observed image, and is used as a test of the code. There are a total of $K=\sum\limits_{i=1,J}K_i=\sum\limits_{i=1,I}(J_i)^2$ relensed images in the whole cluster. 

There are three possible ways to calculate the lens plane rms using all $K$ individual backprojected images. The first is to sum up the deviations of all the $K$ model predicted (i.e., relensed) images from the observed images, in quadrature. Of all five methods we present here, this is the most conservative calculation because it explicitly takes into account every model-predicted image, and adds them in quadrature. It typically yields the largest rms value.  
\begin{equation}
\Big(\Delta^{\rm I}_{\rm rms,tot}\Big)^2=\frac{1}{K}\sum\limits_{i=1,I}
\Bigl\{\sum\limits_{j=1,J_i}\Bigl[\sum\limits_{k=1,J_i}\Big|\,\vec\theta_{i,j,k}-\vec\theta_{i,j}\Big|^2\Bigr]\Bigr\}.
\label{rmstot}
\end{equation}
While in eq.~\ref{rmstot} sources with more images contribute more to the rms, in the second method all sources contribute equally, regardless of the number of images they have,
\begin{equation}
\Big(\Delta^{\rm I}_{\rm rms,src}\Big)^2=\frac{1}{J}\sum\limits_{i=1,I}
\Bigl\{\sum\limits_{j=1,J_i}\Bigl[\frac{1}{J_i}\sum\limits_{k=1,J_i}\Big|\,\vec\theta_{i,j,k}-\vec\theta_{i,j}\Big|^2\Bigr]\Bigr\}.
\label{rmssrc}
\end{equation}
Finally, the average position of the $J_i$ relensed images is compared to the corresponding observed image position. The distance between these are 
\begin{equation}
|\vec\theta_{\rm obs}-\vec\theta_{\rm mod~I}|=(\Delta\theta_{i,j;x}^2+ \Delta\theta_{i,j;y}^2)^{1/2},
\label{distI}
\end{equation}
where
\begin{equation}
\Delta\theta_{i,j;x}=\Bigl(\frac{1}{J_i}\sum\limits_{k=1,J_i}\theta_{i,j,k;x}\Bigr)-\theta_{i,j;x},
\end{equation}
and similarly for the $y$-component. These distances are summed in quadrature to produce,
\begin{equation}
\Big(\Delta^{\rm I}_{\rm rms,ims}\Big)^2=\frac{1}{J}\sum\limits_{i=1,I}\Big\{\sum\limits_{j=1,J_i}
\Big|\,\vec\theta_{\rm obs}-\vec\theta_{\rm mod~I}\Big|^2\Big\}
\label{rmsims}
\end{equation}
Typically, eq.~\ref{rmsims} yields the smallest of eq.~\ref{rmstot}, eq.~\ref{rmssrc}, and eq.~\ref{rmsims} estimated rms values.

(II) Another way to calculate the lens plane rms is to first find the average of the $J_i$ model backprojected images in the source plane. This gives the model-predicted source position. This is then lensed forward to obtain $J_i$ model predicted images, one per observed image. The distance between these corresponding model-predicted and observed images is calculated,
\begin{equation}
\Big|\vec\theta_{\rm obs}-\vec\theta_{\rm mod~II}\Big|^2=
(\theta_{{\rm mod~II};~i,j;x}-\theta_{i,j;x})^2+(\theta_{{\rm mod~II};~i,j;y}-\theta_{i,j;y})^2
\label{distII}
\end{equation}
and used as in eq.~\ref{rmsavg}, except that eq.~\ref{distII} is different from eq.~\ref{distI}. 
\begin{equation}
\Big(\Delta^{\rm II}_{\rm rms,ims}\Big)^2=\frac{1}{J}\sum\limits_{i=1,I}\Big\{\sum\limits_{j=1,J_i}
\Big|\,\vec\theta_{\rm obs}-\vec\theta_{\rm mod~II}\Big|^2\Big\}
\label{rmsimsII}
\end{equation}
We believe that most papers use this definition,
however, some, like \cite{dal11} take the average of eq.~\ref{distII} values, one per source, instead of adding them in quadrature:
\begin{equation}
\Delta^{\rm II}_{\rm rms,avg}=\frac{1}{J}\sum\limits_{i=1,I}\Big\{\sum\limits_{j=1,J_i}
\Big|\,\vec\theta_{\rm obs}-\vec\theta_{\rm mod~II}\Big|\Big\}
\label{rmsavg}
\end{equation}

The values of all five rms estimators (eq.~\ref{rmstot}, eq.~\ref{rmssrc}, eq.~\ref{rmsims}, eq.~\ref{rmsimsII} and eq.~\ref{rmsavg}) for the four reconstructions presented in this paper are shown in Table~\ref{table1}. 
The distributions (per image) are shown in Fig.~\ref{LPrms}, for three estimators only: eq.~\ref{rmstot}, eq.~\ref{rmsims} and eq.~\ref{rmsimsII}.

\bsp	
\label{lastpage}
\end{document}